\documentclass[final,5p,times,twocolumn]{elsarticle}
\usepackage{amssymb}
\usepackage{amsmath}
\usepackage{graphicx}
\usepackage{dcolumn}
\usepackage{bm}
\usepackage[separate-uncertainty=true]{siunitx}
\usepackage{braket}
\usepackage{hyperref}

\usepackage{ulem}
\usepackage[dvipsnames]{xcolor}
\newcommand{\red}[1]{\textcolor{black}{#1}}

\journal{Nucl. Instrum. Methods Phys. Res. A}

 \hypersetup{draft} 

\begin{document}

\begin{frontmatter}

\title{A Facility for Production and Laser Cooling of Cesium Isotopes and Isomers}

\author[label1]{Alexandros Giatzoglou}
\author[label1]{Tanapoom Poomaradee}
\author[label2]{Ilkka Pohjalainen}
\author[label2]{Sami Rinta-Antila}
\author[label2]{Iain D. Moore}
\author[label3]{Philip M. Walker}
\author[label1]{\\Luca Marmugi}
\ead{l.marmugi@ucl.ac.uk}
\author[label1]{Ferruccio Renzoni}
\address[label1]{Department of Physics and Astronomy, University College London, Gower Street, London WC1E 6BT, United Kingdom}
\address[label2]{Department of Physics, University of Jyv\"askyl\"a, Survontie 9, 40014  Jyv\"askyl\"a, Finland}
\address[label3]{Department of Physics, University of Surrey, Guilford GU2 7XH, United Kingdom}

\begin{abstract}
We report on the design, installation, and test of an experimental facility for the production of ultra-cold atomic isotopes and isomers of cesium. The setup covers a broad span of mass numbers and nuclear isomers, allowing one to directly compare chains of isotopes and isotope/isomer pairs. Cesium nuclei are produced by fission or fusion-evaporation reactions using primary proton beams from a \SI{130}{\mega\electronvolt} cyclotron impinging upon a suitable target. The species of interest is ejected from the target in ionic form, electrostatically accelerated, mass separated, and routed to a science chamber. Here, ions are neutralized by implantation in a thin foil, and extracted by thermal diffusion. A neutral vapor at room temperature is thus formed and trapped in a magneto-optical trap. Real-time fluorescence imaging and destructive absorption imaging provide information on the number of trapped atoms, their density, and their temperature. Tests with a dedicated beam of $^{133}$Cs$^{+}$ ions at \SI{30}{\kilo\electronvolt} energy confirm neutralization, evaporation, and laser cooling to \SI{150}{\micro\kelvin}, with an average atomic density of \SI{e10}{\per\centi\metre\cubed}. Availability of cold and dense atomic samples of Cs isotopes and isomers opens new avenues for high-precision measurements of isotopic and isomeric shifts thereby gaining deeper insight into the nuclear structure, as well as for sensitive measurements of isotopes' concentration ratios in trace quantities. The facility also constitutes the core for future experiments of many-body physics with nuclear isomers.
\end{abstract}


\end{frontmatter}

%\linenumbers

\section{Introduction}
Laser spectroscopy of radioactive atoms has the potential to shed new light on open questions about nuclei and fundamental symmetries \cite{kluge2003, palffy2010, cheal2010, blaum2013, campbell2016}. However, progress in this field faces many experimental challenges: in addition to the main problems constantly addressed in atomic and laser physics, the isotopes' short half-life requires dedicated experimental facilities and protocols, at the boundary between nuclear and atomic physics. 

In this context, cold and ultra-cold atoms offer a unique opportunity, allowing one to collect large samples of rare and unstable isotopes, and providing the conditions for high precision laser spectroscopy -- with sub-Doppler linewidths of the order of a few \SI{}{\mega\hertz} -- and long interrogation times -- many orders of magnitude longer than the lifetime of the electronic state excited by lasers \cite{behr2009, willmann2012}.  For example, laser cooling of heavy unstable atoms (mostly Fr \cite{lnlol, orozco2018}) has given access to unprecedented spectroscopic data, motivated by the study of phenomena such as parity non-conservation in atoms and the permanent electronic dipole moment \cite{willmann2012, gwinner2006, sakemi2011, mariotti2014}.

In this paper, we report on the facility for laser cooling of cesium isotopes and isomers built at the Accelerator Laboratory of the University of Jyv\"askyl\"a (Finland). From the production site of the ions of interest to the trap of cold atoms, the energy scale spans from \SI{e4}{\electronvolt} (\SI{e6}{\electronvolt} in the case of the primary proton beam) to \SI{e-8}{\electronvolt} in around \SI{5}{\second}. The facility, the first of its kind entirely dedicated to Cs, is designed to trap wide mass range of Cs isotopes (from A=133 to 142), in their nuclear ground and long-lived excited states (A=134m, 135m, 136m, 138m). This opens up new avenues for the investigation of charge radii anomalies in $^{\text{A,Am}}$Cs pairs \cite{bissel2007, dracoulis2016}, and for the creation and validation of nuclear forensics techniques and instrumentation \cite{dirosa2003, yang2016}. The ability to manipulate ultra-cold radioactive atoms will also allow the exploration of yet-to-be observed many-body effects \cite{marmugi2018}.

The setup can be ideally divided into a high-energy section (production, acceleration, and mass separation of ions) and a low-energy section (thermalization, neutralization, and laser cooling and trapping). The description of the experimental facility presented here follows the order of operation: in Sec.~\ref{sec:overview} an overview of the facility is presented. In Sec.~\ref{sec:highenergy} a detailed description of the high energy section of the experimental setup is provided, while in Sec.~\ref{sec:lowenergy} the low energy (``cold atoms'') section is described. Finally, in Sec.~\ref{sec:experiment}, a full experimental characterization of the facility with a $^{133}$Cs$^{+}$  beam is reported. Results obtained during implantation tests with $^{135m,135}$Cs$^{+}$ are also reported.

\begin{figure*}[ht]
\begin{center}
\includegraphics[width=\linewidth]{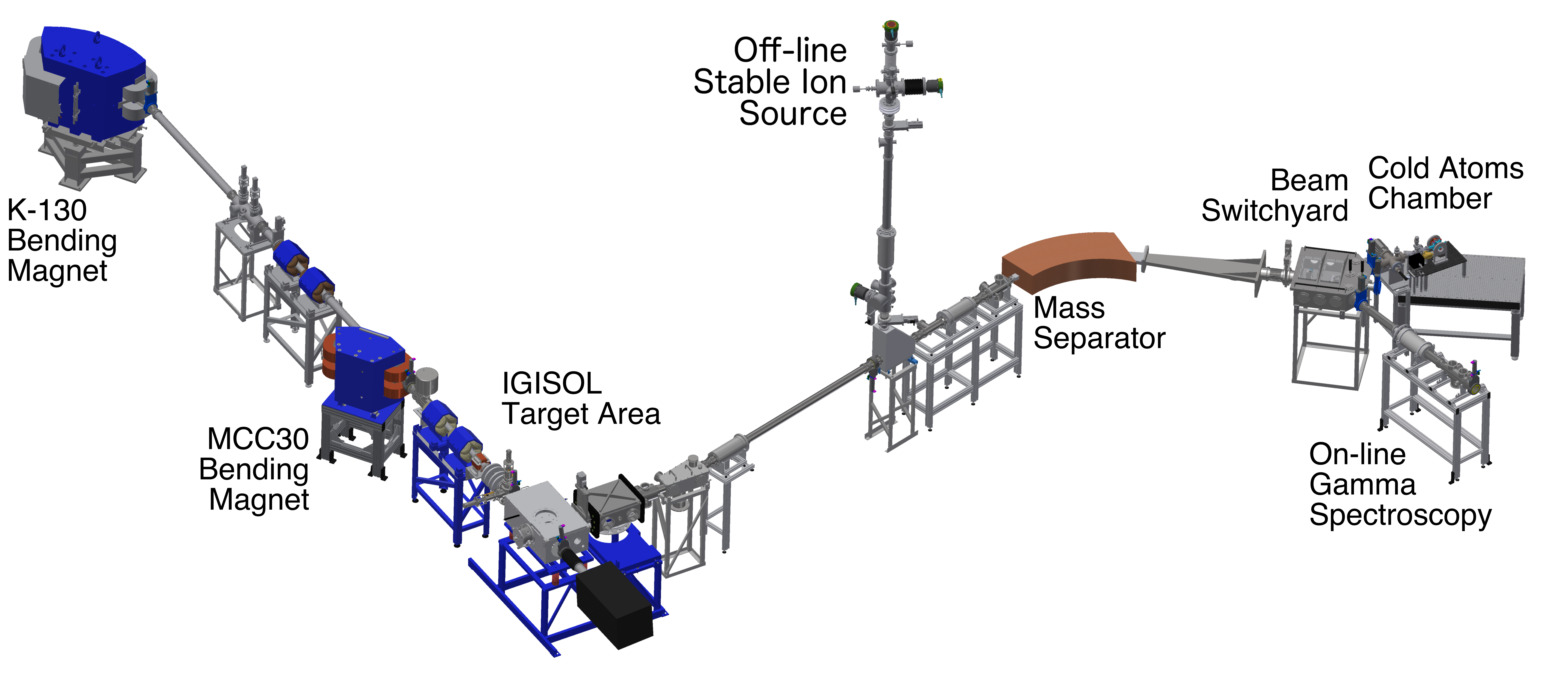}
\caption{(Color online) Overview of the IGISOL facility. The vacuum chamber of the cold atoms experiment is visible on the right hand side. The optical and laser set-up is not shown for clarity. Details are in the text.}
\end{center}
\label{fig:setup}
\end{figure*} 

\section{Overview of the Facility}\label{sec:overview}
An overview of the experimental setup, comprising the Ion Guide Isotope Separation On-Line (IGISOL) mass separator, an off-line source of $^{133}$Cs$^{+}$, and the low energy section, where laser cooling is performed, is presented in Fig.~\ref{fig:setup}.

The primary beam is produced in the K130 cyclotron of the Accelerator Laboratory of the University of Jyv\"askyl\"a. Fission \red{(e.g. in the case of $^{135,135m}$Cs, discussed in Sec.~\ref{subsec:radioactive}) or fusion-evaporation (e.g. in the case of $^{134,134m}$Cs, promising for radius anomaly studies \cite{newref})} products following nuclear reactions are electrostatically accelerated towards the low-energy sector. Here, a dedicated $\gamma$ spectroscopy line -- within the IGISOL facility -- can be used for monitoring the isomeric content of the beam. Alternatively, the ionic beam is routed towards the science chamber, where ions are neutralized via thin foil implantation, extracted as an atomic vapor, and trapped in a Magneto-Optical Trap (MOT). The atomic sample -- cooled down to \SI{150}{\micro\kelvin} -- is monitored in real-time with fluorescence imaging, or with absorption imaging with a repetition rate of \SI{0.1}{\hertz}. For testing the operation of the system, and retaining the possibility of direct comparison with the stable reference isotope, an off-line source of $^{133}$Cs$^{+}$ was realized on the second floor of the facility (top of Fig.~\ref{fig:setup}). The ions from the off-line source, once merged into the main separator, share the same path with the radioactive beam. This also allows testing and optimisation of the transport line.

\section{High-Energy Section}\label{sec:highenergy}
At IGISOL, nuclear reactions are induced using either light ions (protons, deuterons or alpha particles) or heavy ions from either a new MCC30 \red{light-ion} cyclotron, or a K130 cyclotron \cite{igisol1}. In this work, the K130 cyclotron is used, with a \SI{50}{\mega\electronvolt} proton beam impinging on a natural U target in order to induce fission. The fission products recoil isotropically from the target into a buffer gas cell located in the IGISOL target area (Fig.~\ref{fig:setup}) and operating with helium at typical pressures of \SI{300}{\milli\bar}. Approximately 1\% of the fragments are thermalized and stopped in the gas \cite{igisol2}, with the majority of isotopes surviving as singly-charged ions. The reaction products, along with the buffer gas and any impurities, are swept out of the gas cell through a \SI{1.2}{\milli\meter} diameter exit hole, whereby the neutral gas is pumped away using high-throughput vacuum pumps.

The ions are subsequently captured in the radiofrequency field of a sextupole ion guide (SPIG) \cite{igisol3} and are guided towards the high vacuum region of the mass separator. At this stage, the ions are accelerated to a potential of \SI{30}{\kilo\electronvolt}, electrostatically transported, and mass separated using a \SI{55}{\degree} dipole magnet with a typical mass resolving power $300\leq M/\Delta M \leq 500$.
%, and are detected using a variety of diagnostics available at the focal plane of the separator.}

\subsection{Off-line Source of $^{133}$Cs$^{+}$}
\begin{figure}[htbp]
\begin{center}
\includegraphics[height=6cm]{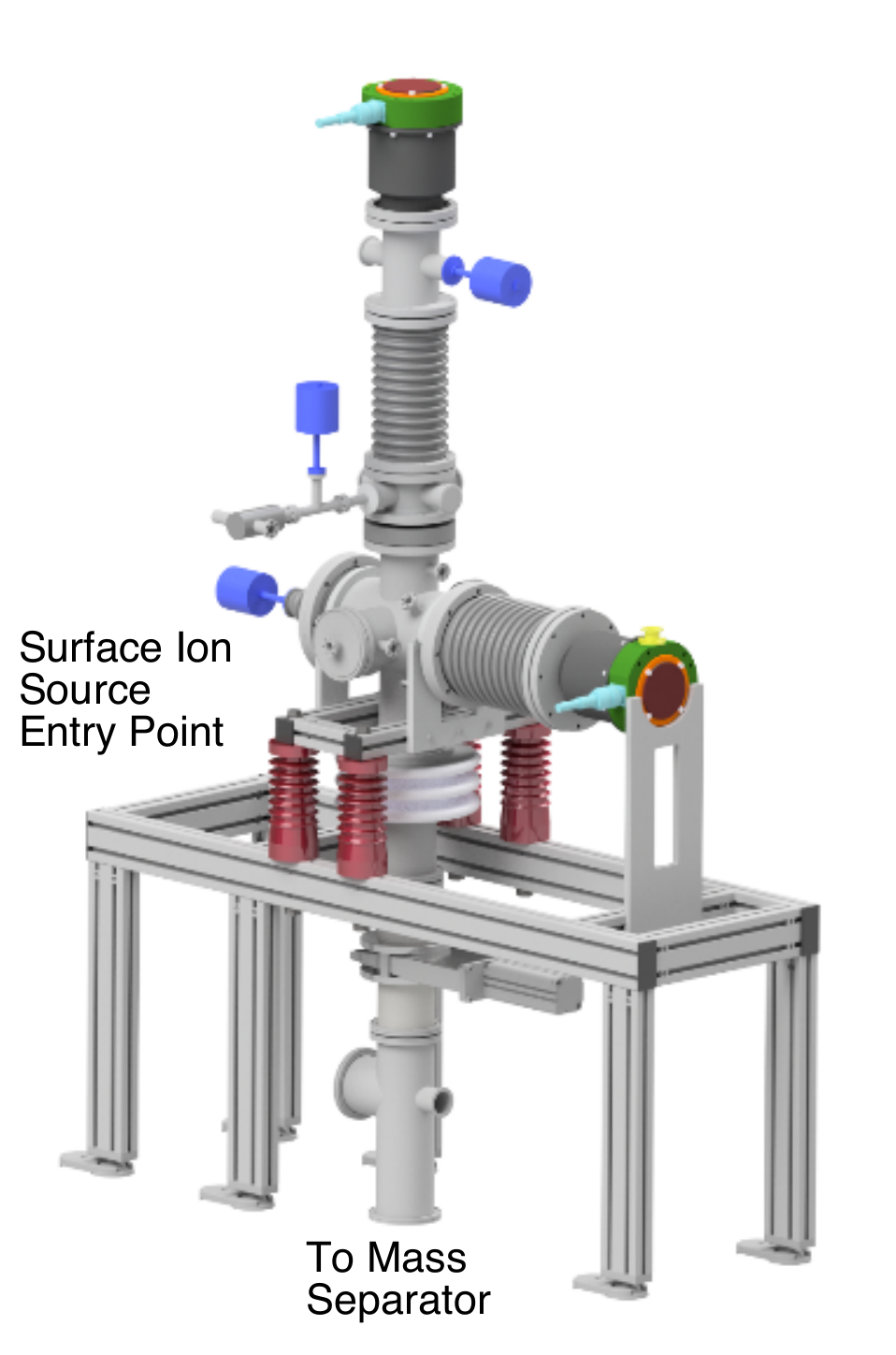}
\caption{(Color online) Stable ion source (``off-line source''). A surface ion source with the desired species -- in the present case $^{133}$Cs -- is inserted on the left hand side. Ionization of Cs and other alkali elements occurs via the hot surface of a filament. The ions are then electrostastically accelerated by a \SI{30}{\kilo\volt} electrode and are routed to the IGISOL mass separator.\label{fig:localsource}}
\end{center}
\end{figure}

Rather recently, a new off-line ion source was constructed and installed in order to provide stable beams of ions from a location on the second floor of the IGISOL facility (Fig.~\ref{fig:localsource}).

The use of the off-line source allows for a more flexible and safe operation, with off-line operation available immediately following an on-line experiment without the need to wait for a ``cooling'' period. It also enables a stable reference source to be provided in-situ with radioactive beams from the front-end of the facility. Two ion sources are in current operation, a surface ion source and a so-called discharge ion source. The latter has been presented recently in connection with the production of an ion beam of stable $^{89}$Y for collinear laser spectroscopy \cite{igisol4}.

The surface ion source is used for the production of a $^{133}$Cs$^{+}$ beam. Ions are accelerated using a dedicated high voltage platform to the same operating potential as the front-end of the IGISOL, namely \SI{30}{\kilo\volt}. Such a beam of ions is mass separated in an identical manner to the on-line operation and transported to the science chamber of the low-energy section for further manipulation.

\section{Low-Energy Section}\label{sec:lowenergy}
The low-energy section is located immediately downstream from the focal plane of the IGISOL mass separator, occupying a new beam line, highlighted in Fig.~\ref{fig:setup}. This section is differentially pumped, and isolated by an ultra-high vacuum (UHV) pneumatic gate valve after the IGISOL switchyard. A \SI{350}{\litre\per\second} turbo-molecular pump (Leybold CE TURBOVAC 350i), ensures a stable vacuum $\leq\SI{e-8}{\milli\bar}$, even when the gate valve is abruptly opened. A custom-made CF63 cross element secures the turbo-molecular pump to the main transport line, provides support for the science chamber, and allocates diagnostic tools such a movable Faraday cup, and a Penning gauge. The cross element is connected to a CF63/CF40 adaptor, which in turn contains a second vacuum impedance.

As shown in Fig.~\ref{fig:cell}, an ion pump (\SI{20}{\litre\per\second}, Varian VacIon Plus 20 StarCell) produces a residual pressure in normal operation of $\SI{1e-9}{\milli\bar}$. The ion pump is isolated from the experimental chamber during maintenance by a second UHV gate valve. We have observed that -- together with venting with N$_{2}$ -- this dramatically reduces the waiting time for breaking and re-establishing operational vacuum conditions. A few hours, as opposed to two days, are usually enough, allowing emergency maintenance to be performed. \red{The ion pump's residual magnetic field at the beam's position has been taken into account during transport optimization. Accordingly, no mu-metal shielding is used.}

A CF40 cross provides connection to a CF40 valve, used for pre-evacuation with a removable \SI{50}{\litre\per\second} turbo-molecular pump and for N$_{2}$ venting. On the opposite side, a CF16 valve is installed and connected to a blind CF16 flange, where a reservoir of $^{133}$Cs has been sealed. Atoms are extracted -- when needed -- via thermal evaporation, and are usually detected in the main experimental chamber within a few seconds after the valve has been opened. Cs vapor provides another tool for testing the optical setup and the operation of the MOT. Attached to the remaining port of the CF40 cross is the experimental chamber, whose main axis coincides with that of the ion transport line.

\subsection{Science Chamber}
\begin{figure}[htbp]
\begin{center}
\includegraphics[height=6cm]{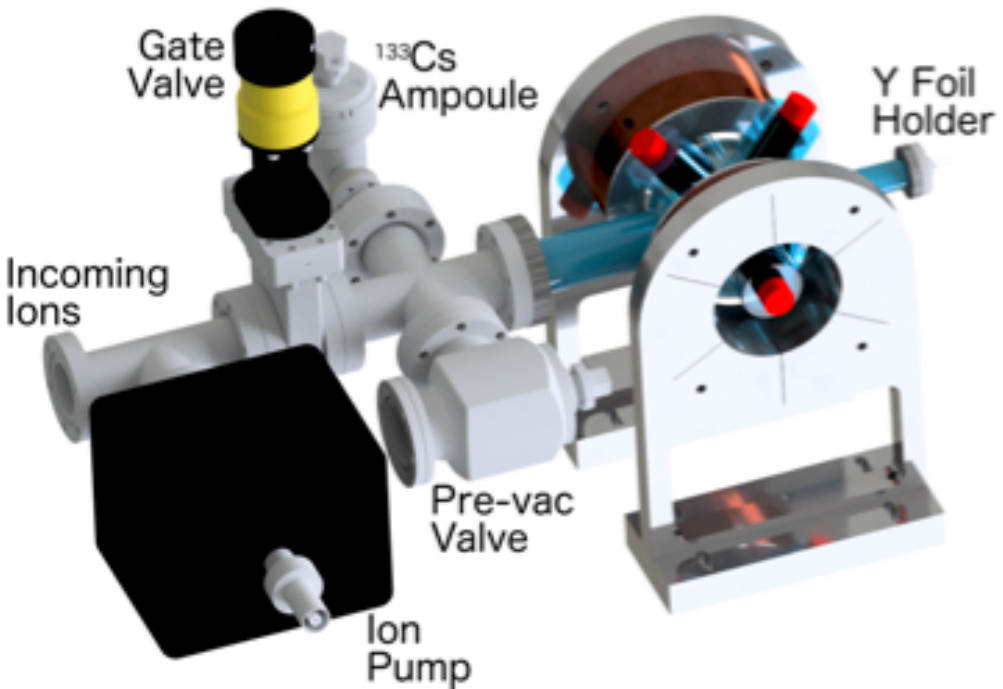}
\end{center}
\caption{(Color online) Rendering - to scale - of the terminal part of the low energy section. The borosilicate glass science chamber is shown on the right hand side. The ion beam enters from the CF40 connection on the left, next to the \SI{20}{\litre\per\second} ion pump, passes through the gate valve, and enters the laser cooling chamber. A CF16 blind flange equipped with electrical feedthroughs supports an implantation foil. Anti-Helmholtz quadrupole coils are visible at either side of the glass chamber. \red{The implantation thin foil (not visible) is placed \SI{20}{\milli\meter} inside the science chamber. Further details are shown in Fig.~\ref{fig:neutralizer}.} Positions of the laser beams are marked by red cylinders. A solid reservoir of $^{133}$Cs is sealed by the CF16 valve visible in the background.}
\label{fig:cell}
\end{figure} 

The science chamber is a borosilicate glass spheroid, equipped with \SI{40}{\milli\metre} diameter, 40-20 scratch-dig optical windows, supported by dedicated borosilicate glass tubes, and enclosed by two coils (Fig.~\ref{fig:cell}). The chamber inner diameter is around \SI{100}{\milli\metre}. The overall length is \SI{250}{\milli\metre}, including the connection to the vacuum line and the connection to the neutralizer feedthrough as described below. Optical windows are aligned in pairs to allow easy alignment of the laser beams for the MOT. Two are placed along the axis orthogonal to the transport line, along the main axis of the main coils. The other two pairs of optical windows are located in the plane parallel to the coils' faces with a separation of around \SI{80}{\degree}.

Connection to the vacuum line is ensured by a CF40 metal-to-glass connector. This also acts as a mechanical support for the whole chamber. Extra support is provided by tailored foam segments underneath the two ``horizontal'' windows resting on the main coils mounts (not shown in Fig.~\ref{fig:cell}). This also dampens mechanical vibrations, without producing sheer stress that could damage the chamber in the long term.

The two main coils are in the ideal anti-Helmholtz configuration, generating a magnetic quadrupole field. The system is air-cooled and creates a linear magnetic field gradient of \SI{10}{\text{G}\per\centi\metre\per\ampere} in the trapping region. Current in the coils is controlled by a water-cooled MOSFET (IXFN 120N20), driven by an SRS SIM960 PID controller. The setpoint of the PID loop is controlled via LabVIEW, and can be changed during the experimental sequence via the control software \cite{nolli2016}. In addition, three pairs of mutually orthogonal squared compensation coils (not shown in Fig.~\ref{fig:cell}) nullify the background DC fields acting on the atoms, and control the position of the atomic trap.

\subsubsection{Anti-relaxation Organic Coating}
The inner surface of the experimental chamber is coated with a layer of polydimethylsiloxane (PDMS, CH$_{3}$[Si(CH$_{3}$)$_{2}$O]$_{n}$Si(CH$_{3}$)$_{3}$). The coating is a strategy for minimizing atomic losses for high efficiency laser cooling experimental \red{chambers \cite{stephens1994, lu1997, scireplnl}}. In detail, PDMS lowers the glass surface adsorption energy down to $\SI{\sim0.1}{\electronvolt}$, drastically reducing chemisorption after atom/wall collisions. Consequently, the amount of atoms of interest available in the experimental chamber is increased. This is a paramount feature in the case of short-lived species, which allows the cold atom MOTs to be maximally populated.

The chosen PDMS (Sigma-Aldrich) has a viscosity of \SI{65}{\centi\text{St}}. For deposition, we have adopted a protocol based on a modified version of the procedure presented in \cite{scireplnl}. The glass is prepared in a prolonged bath of 0.45:0.45:0.10 volume solution of CH$_{3}$OH, C$_{2}$H$_{5}$OH, and KOH. KOH is slowly and carefully dissolved in the liquid solution with gentle stirring, making sure that the mixture's temperature does not exceed \SI{40}{\celsius}. The chamber is rinsed five times with de-ionized water, and then is left in a ventilated chemical hood at room temperature for \SI{30}{\minute}. The chamber is then connected to a turbo-molecular pump (\SI{50}{\litre\per\second}). Once a stable pressure of \SI{1e-6}{\milli\bar} is reached, gentle baking with resistive heating tape is performed. Three layers of Al foil isolate the system from air and ensure uniform heating of the chamber. The temperature is increased at a rate of \SI{10}{\celsius}/\SI{20}{\minute}, until  \SI{120}{\celsius}. The procedure is terminated when the pressure of \SI{1e-7}{\milli\bar} is stable. Cooling of the chamber is obtained with spontaneous thermalization, after switching off the heating tape.

The coating is applied by spin-coating (\SI{0.1}{\text{rps}}) of a solution 0.05:0.95 of PDMS and (C$_{2}$H$_{5}$)$_{2}$O. After five minutes, the spin is interrupted and the excess solution is eliminated. The chamber is then baked up to \SI{100}{\celsius}, at a rate of \SI{10}{\celsius}/\SI{20}{\minute}, until the baseline pressure of \SI{1e-7}{\milli\bar} is reached. After installation in the main vacuum line, baking of the whole system is performed. A baseline pressure of \SI{1e-9}{\milli\bar} is usually obtained within \SI{24}{\hour}. A final curing (``passivation'') is obtained by using the thermal Cs vapor from the solid Cs reservoir. The progression of the curing process is monitored by observing the stability of the number of atoms trapped in the MOT obtained from thermal vapor.

\subsection{Ion Neutralization and Atom Extraction}
\begin{figure}[htbp]
\begin{center}
\includegraphics[height=6cm]{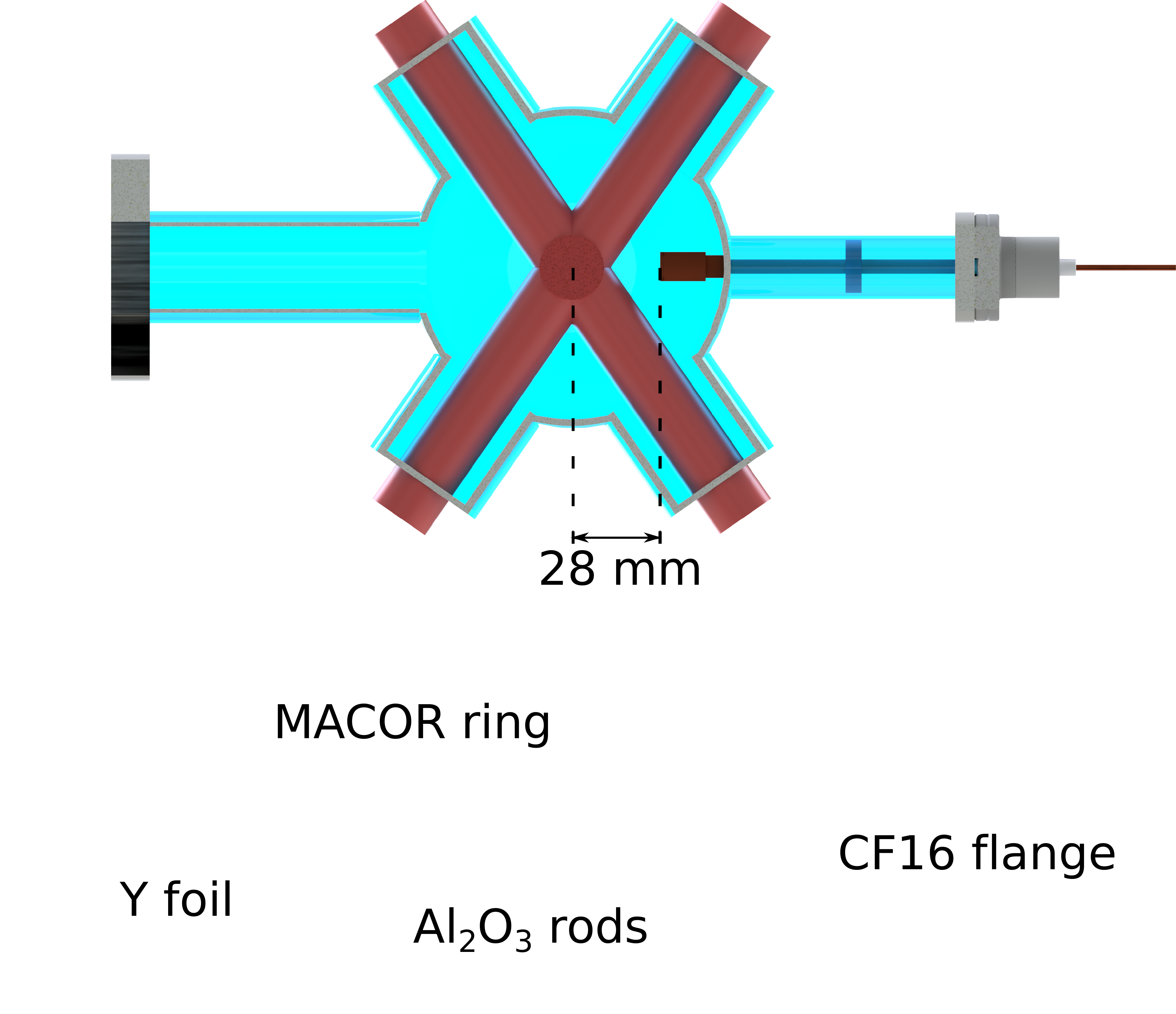}
\end{center}
\caption{(Color online) Sketch - to scale - of the neutralizer mount. A thin Y foil, where the incoming ion beam is focused, is supported by copper contacts, in turn connected to electrical feedthroughs. Mechanical stability is ensured by ceramic rods, secured to a CF16 flange. Resistive heating allows one to control the temperature of the foil, up to \SI{e3}{\kelvin}. \red{Top: relative position of the Y foil and the MOT (approximately at the center of the laser beams overlapping region). Bottom: detailed view of the neutralizer mount. }}
\label{fig:neutralizer}
\end{figure}

The incoming ion beam is slowed down and neutralized via thin foil \red{implantation \cite{dirosa2003,gwinner1994}.} Ions are electrostatically  focused on a \SI{25}{\micro\meter} thick yttrium foil ($\SI{16}{\milli\metre}\times\SI{9}{\milli\metre}$), held approximately \SI{20}{\milli\meter} inside the science chamber \red{(for details and the actual position, see Figs.~\ref{fig:neutralizer})}. 

The Y foil (99.9$\%$ REO, supplied by Alfa Aesar) is kept in place by copper supports, connected to two hollow aluminum oxide (Al$_{2}$O$_{3}$) rods, \SI{105.6}{\milli\metre} long. The rods are mechanically secured to the \SI{20}{\ampere}-rated electrical feedthroughs of a CF16 flange. Electrical connections are ensured by a hollow Cu core embedded in the ceramic rods, connecting \SI{1.5}{\milli\metre} diameter Cu wire of the feedthroughs to the supports of the Y foil. With a newly installed foil, the circuit has a resistance of $1.2\pm\SI{0.1}{\ohm}$, where the uncertainty takes into account the statistical variability observed. A MACOR centering ring, of \SI{16.5}{\milli\meter} diameter, ensures the correct positioning of the supporting rods and prevents excess tension on the thin Y foil. At the same time, mechanical tolerance allows free expansion of the foil upon increase of the temperature.

\begin{figure}[h]
\begin{center}
\includegraphics[height=6.5cm]{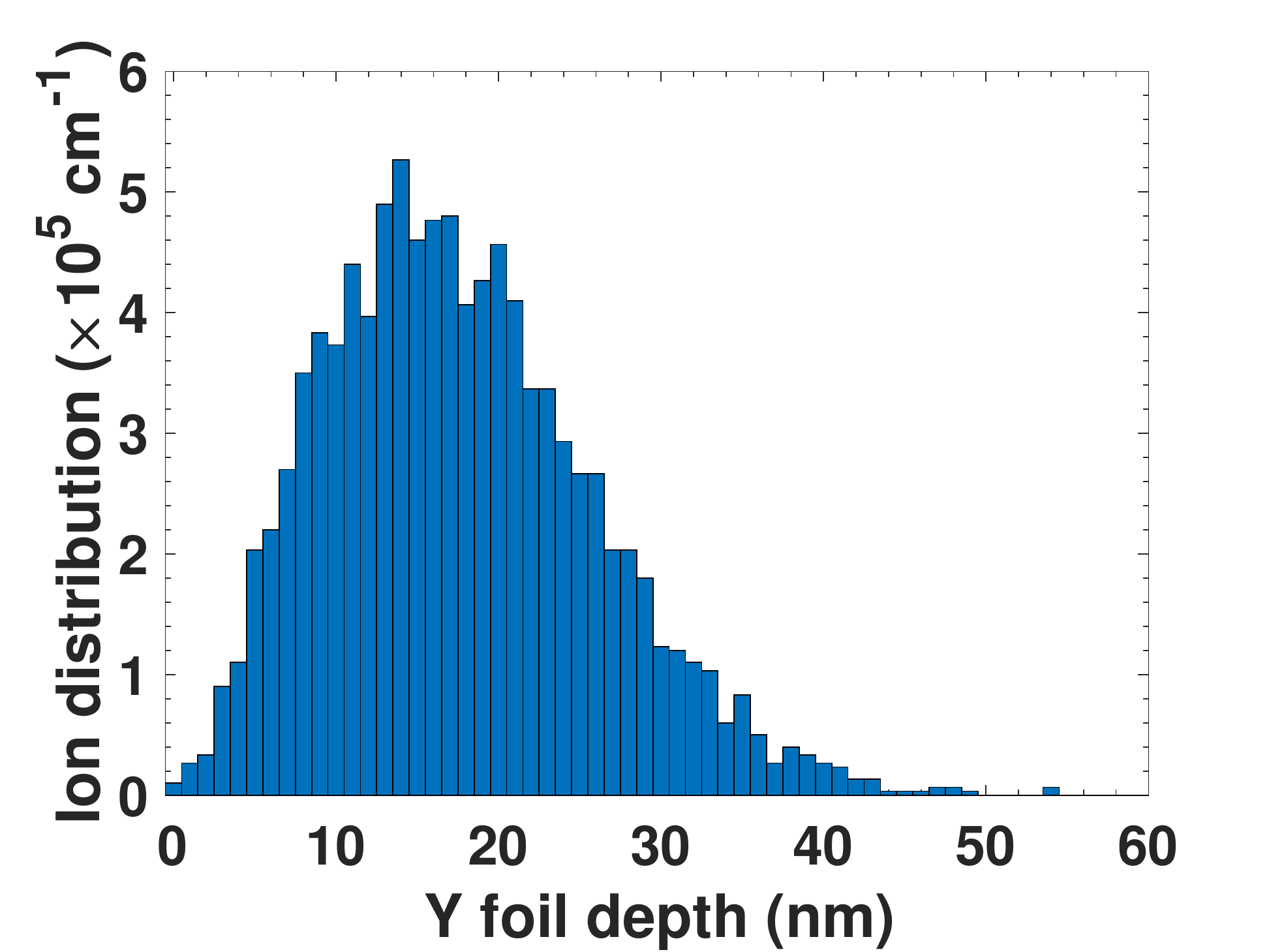}
\end{center}
\caption{(Color online) SRIM simulations of the implantation of  \SI{30}{\kilo\electronvolt} $^{133}$Cs$^{+}$ in the \SI{25}{\micro\metre} thick Y foil: linear density of the implanted ions. For this simulation, 10$^{4}$ ions were used, with a target length of \SI{100}{\nano\meter}.}
\label{fig:implantation}
\end{figure}

SRIM simulations \cite{srim} indicate a peak implantation density around \SI{150}{\angstrom} from the surface of the foil for the \SI{30}{\kilo\electronvolt} ions (Fig.~\ref{fig:implantation}).  With respect to recent works with Fr \cite{scireplnl}, the implantation depth is significantly larger, causing a greater involvement of the bulk of the foil in the storage of the incoming ions. This configuration is more efficient in the case of long-lived particles, such as those of interest here.

%%% Ion input temperature associated to motion
 %(\SI{3.5e8}{\kelvin}) to \SI{e-1}{\electronvolt} (\SI{e3}{\kelvin})

Implanted ions dissipate a large fraction of their kinetic energy via inelastic collisions with the metal lattice, and slow down to a diffusive regime. Because of the favorable imbalance between the Cs ionization potential (I$_{\text{Cs}}$=\SI{3.89}{\electronvolt} \cite{weber1987}) and the Y work function (W$_{\text{Y}}$=\SI{3.1}{\electronvolt}\red{, as estimated for a pure Y sample}), they also acquire an extra electron from the surrounding metal, thus achieving neutralization required for laser cooling. Inside the foil, the relative concentration of neutral atoms ($n$) and ions ($n_{+}$) is given by the Langmuir-Saha equation \cite{jchemphyslnl}:

\begin{equation}
\dfrac{n_{+}}{n}=\dfrac{g_{+}}{g}\exp\left( \dfrac{W_{Y}-I_{Cs}}{k_{B}T}\right)~,\label{eqn:saha}
\end{equation} 

\noindent where $k_{B}$ is the Boltzmann constant, $T$ is the foil's temperature in \SI{}{\kelvin}, and $g_{+}/g=1/2$. In operational conditions, $n_{+}/n\leq10^{-4}$. \red{The value of W$_{\text{Y}}$ used in Eq.~\ref{eqn:saha} may also vary with the surface’s lattice structure and contamination. We note that an alternative candidate for the neutralizer foil is Zr, as recently adopted in Fr experiments with promising results \cite{zr}.}

Atoms are extracted by thermal diffusion, enhanced by heating the foil up to \SI{e3}{\kelvin}, via controlled flow of a DC current (I$_{\text{DC}}$). An optical pyrometer (METIS M3 SensorTherm) is used for measuring the characteristic curves T$_{\text{Y}}$-I$_{\text{DC}}$. We have observed a large variability in the Y foil response with a given DC power supplied, up to a factor 2. We attribute this to the intrinsic purity of the sample: although quantitative analysis was not possible, the nominally most pure foils require a larger current for reaching a given temperature. This is consistent with the presence of fewer impurities that act as scattering centres for the flowing charge carriers. \red{To reduce the impact of contamination during installation, the Y foil is kept in the original Ar atmosphere as long as possible. It is then extracted in air and installed on the dedicated mount in a clean environment at atmospheric pressure. Afterwards, the foil is immediately installed in the science chamber, which has been in the meantime vented with N$_{2}$. Evacuation is immediately started. As soon as stable vacuum is achieved, progressive heating is performed to enhance controlled outgassing of the freshly installed Y foil. The actual rate of heating may vary depending on the vacuum conditions, which are constantly monitored.}
 
Upon thermal desorption in the science chamber, the atoms' velocity distribution is determined by the high temperature required for efficient desorption (\SI{600}{\kelvin}$\leq$T$\leq$\SI{1000}{\kelvin}). Atoms released in this way are too fast to be efficiently laser cooled, given the finite capture range of the cooling mechanism. The PDMS coating allows atoms to thermalize at room temperature: atoms are desorbed with a thermal velocity distribution compatible with room temperature. In this way, the percentage of atoms with velocities below the MOT capture velocity is increased \cite{stephens1994}.

\subsection{$^{133}$Cs MOT Configuration and Laser System}
The MOT is formed by three retro-reflected optical beams, to increase the compactness. Cooling and repumper beams are overlapped and then split in three components, each controlled in power by dedicated polarizing beamsplitter cubes and half-waveplates. Beam-expanders set the size of the laser beams to \SI{20}{\milli\meter} (diameter). Circular polarization is independently obtained with dedicated quarter-waveplates, as standard in cold atom experiments.

The cooling laser light is generated by an extended-cavity diode laser (ECDL) equipped with an intra-cavity interference filter (Radiant Dyes NarrowDiode) tuned to the $6^{2}S_{1/2}(F=4)\rightarrow6^{2}P_{3/2}(F'=4\times5)$ cross-over transition of the D$_{2}$ line. A chain of a double-pass (AOM1) and a single-pass (AOM2) acousto-optic modulators shifts the laser frequency to the $6^{2}S_{1/2}(F=4)\rightarrow6^{2}P_{3/2}(F'=5)$ transition. AOM1 allows the operator to control the fine tuning of the laser, while AOM2 provides fast switching of the laser beam during the experimental sequence \cite{nolli2016}. The cooling laser light  is amplified by a tapered amplifier (m2k GmbH TA-0850-2000). A single-mode polarization-maintaining optical fiber acts as a spatial filter, ensuring a TEM$_{00}$ transverse mode. The intensity of each cooling beam in the trapping region is \SI{4.67}{\milli\watt\per\centi\metre\squared},  corresponding to around 4.2I$_{sat}$, where I$_{sat}$=\SI{1.10}{\milli\watt\per\centi\metre\squared} is the saturation intensity of the cycling transition used for laser cooling.

The repumper laser light  is supplied by a second ECDL (Radiant Dyes NarrowDiode), tuned to the $6^{2}S_{1/2}(F=3)\rightarrow6^{2}P_{3/2}(F'=4)$ optical transition. This laser compensates for hyperfine optical pumping losses, as typical in a MOT. The average repumper intensity is \SI{1.1}{\milli\watt\per\centi\metre\squared}. A mechanical shutter controls the beam during the experimental sequence.

Laser frequency locking is obtained via Dichroic Atoms Vapor Laser Lock (DAVLL) \cite{davll}, and custom designed PI controllers. DAVLL provides locking stability in a noisy environment for several hours, independent of possible laser power fluctuations.

\subsection{Cold Atoms Imaging and Diagnostics}
The MOT is monitored with two different approaches, fluorescence and absorption imaging. Absorption imaging relies on a resonant beam ($6^{2}S_{1/2}(F=4)\rightarrow6^{2}P_{3/2}(F'=5)$, \SI{50}{\micro\watt}), intercepting the MOT along the main coils' axis. This beam is independently controlled by an additional AOM (AOM3), driven by the experimental LabVIEW routine. After interaction with the atomic cloud, the beam is imaged by a computer-triggered CCD camera (PCO PixelFly). Absorption imaging -- destructive for the trap -- is chosen for measuring the MOT's size, density, temperature, and lifetime. Temperature measurements are obtained with free-expansion and time-of-flight measurements \cite{nolli2016}.

Fluorescence imaging collects the spontaneous emission from the trapped atoms during laser cooling. A CCD camera (PCO PixelFly), equipped with an interference filter ($\lambda_{c}=\SI{852}{\nano\meter}$, full-width at half maximum FWHM=\SI{10}{\nano\meter}), images the MOT. A LabVIEW programme continuously monitors the MOT population, its distribution in space, and the cooling laser power stability. The software automatically subtracts the scene's background and compensates for laser intensity fluctuations. An ultimate background noise corresponding to $\sim$20 atoms, with a residual fluctuation of only 3\% in \SI{1}{\hour}, has been measured. This number comprises optical, thermal, and shot noise. Fluorescence imaging can monitor the ``real-time'' evolution of the MOT population, and is therefore well-suited for investigation of the trap dynamics, of the contribution of the ionic beam, and for tuning and optimization  \cite{lnlol, scireplnl}. 

\section{Experimental Characterization and Commissioning of the Facility}\label{sec:experiment}
The operation of the facility is characterized in the following. In particular, the first demonstration of MOT \emph{real-time} loading from $^{133}$Cs$^{+}$ is presented as a validation test for the transport and neutralization setup. The relatively high currents achieved with the beam of stable $^{133}$Cs$^{+}$ ($\SI{\sim{e-9}}{\ampere}$) cannot be representative of the low currents achievable with radioactive beams. Nevertheless, this is beyond the scope of the present work. Furthermore, scaling with current can give a first approximated estimation of the system's response with a smaller number of atoms.

\subsection{$^{133}$Cs$^{+}$-loaded MOT}
In the science chamber, $^{133}$Cs MOTs can be obtained from different sources. One method is direct loading from thermal vapor produced by the neutral cesium sample contained in the CF16 reservoir. This corresponds to the ``conventional'' loading process for a MOT. 

\begin{figure}[h]
\begin{center}
\includegraphics[height=6.5cm]{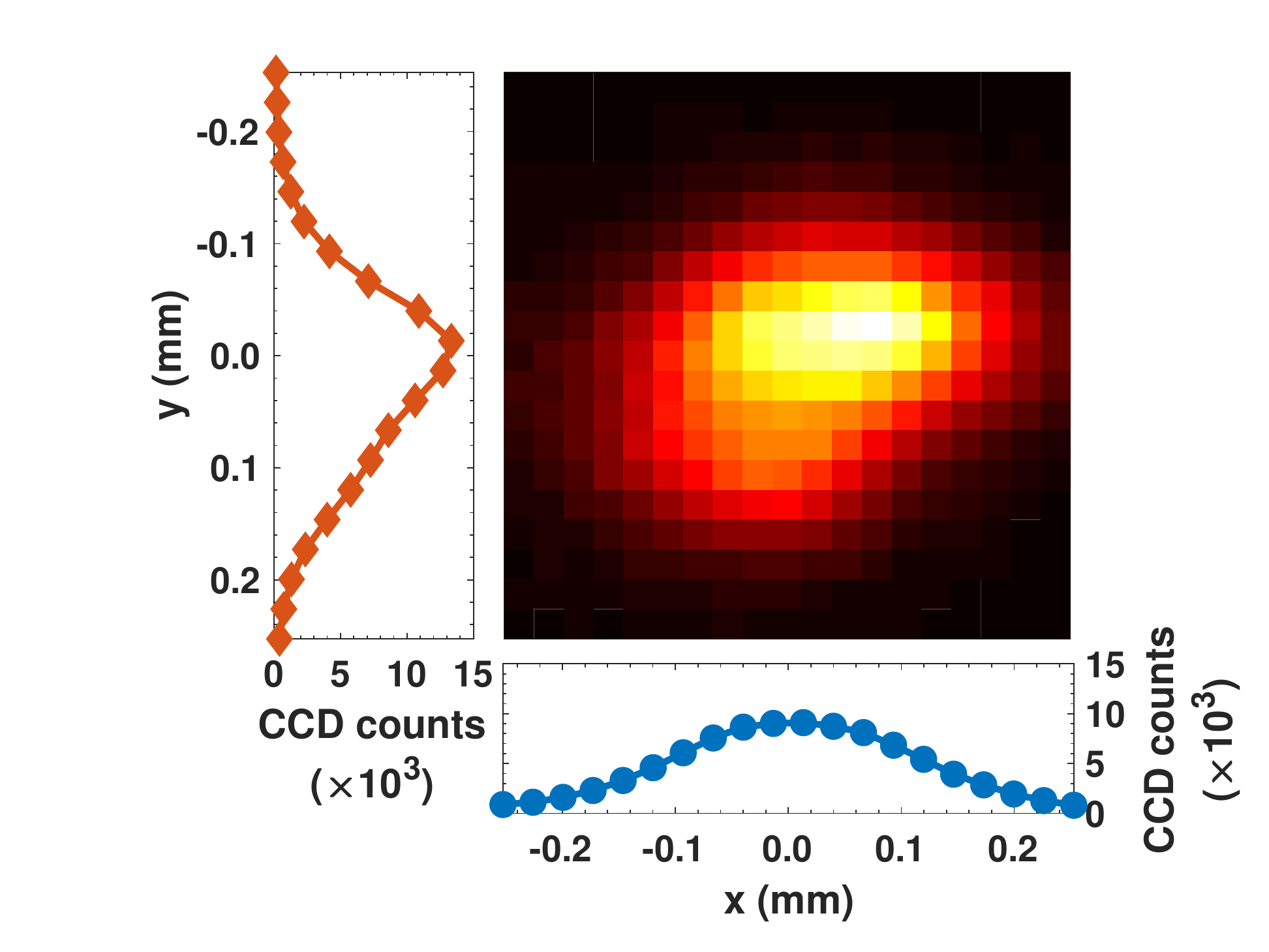}
\end{center}
\caption{(Color online) Typical fluorescence image of a $^{133}$Cs MOT, loaded by heating the Y foil to \SI{931}{\kelvin} (I$_{\text{DC}}$=\SI{5.5}{\ampere}). The trap population is N$_{\text{MOT}}$=1700. The foil has been previously implanted with \SI{1e-10}{\ampere} of \SI{30}{\kilo\electronvolt} $^{133}$Cs$^{+}$. The MOT is therefore loaded in the absence of the $^{133}$Cs$^{+}$ beam. Insets: real time MOT profiles (rows and columns pixel sums). CCD camera exposure is \SI{2}{\second}.}
\label{fig:photo}
\end{figure}

In operational conditions, the residual background partial pressure of Cs vapor is negligible, and no MOT loading can be observed. This condition corresponds to that of radioactive species. In this case, loading can be obtained only via desorption from the Y foil and thermalization on the PDMS coating either from previously implanted neutralized ions, or from freshly implanted neutralized ions. Figure~\ref{fig:photo} shows a fluorescence image -- in false color -- of a $^{133}$Cs MOT,  together with the real time MOT profiles, calculated by the control software. The MOT in Fig.~\ref{fig:photo} is obtained with loading from the Y foil, after implantation with $^{133}$Cs$^{+}$, at an average current of \SI{1e-10}{\ampere}.

% From 2018_04_20_16-45-25_Xavg.txt, 2018_04_20_16-45-25_Yavg.txt
% /Other 133Cs+ data/2018_04_20 - Systematics for low Y temp/Second effort-MOT optimised for low Y temp/Luca - New graphs
% Calibration from 20180619 - Final fluorescence calibration.nb, rescaled for texposure=2 s. Actual value 1709.

No fundamental difference was found among MOTs loaded with different methods. We attribute this to the effectiveness of the neutralization and extraction processes conveyed by the Y foil and the PDMS coating. On average, the optimum cooling laser detuning was found to be $\Delta=-2.65\Gamma$, where $\Gamma=2 \pi \times\SI{5.2}{\mega\hertz}$ is the $F=4\rightarrow F'=5$ transition's natural linewidth. The typical MOT density is \SI{e10}{\per \centi \metre \cubed}, with a temperature T=($150 \pm 10)~\SI{}{\micro\kelvin}$. The uncertainty takes into account instrumental uncertainties and statistical fluctuations. A very small asymmetry ($\pm\SI{1}{\micro\kelvin}$) was measured along the two orthogonal axes, because of a very good expansion symmetry. This confirms effective compensation of the background fields and proper alignment of the imaging system. We note that -- as expected -- the MOT temperature can significantly vary, depending on the chosen parameters.

\begin{figure}[htbp]
\begin{center}
\includegraphics[height=6.5cm]{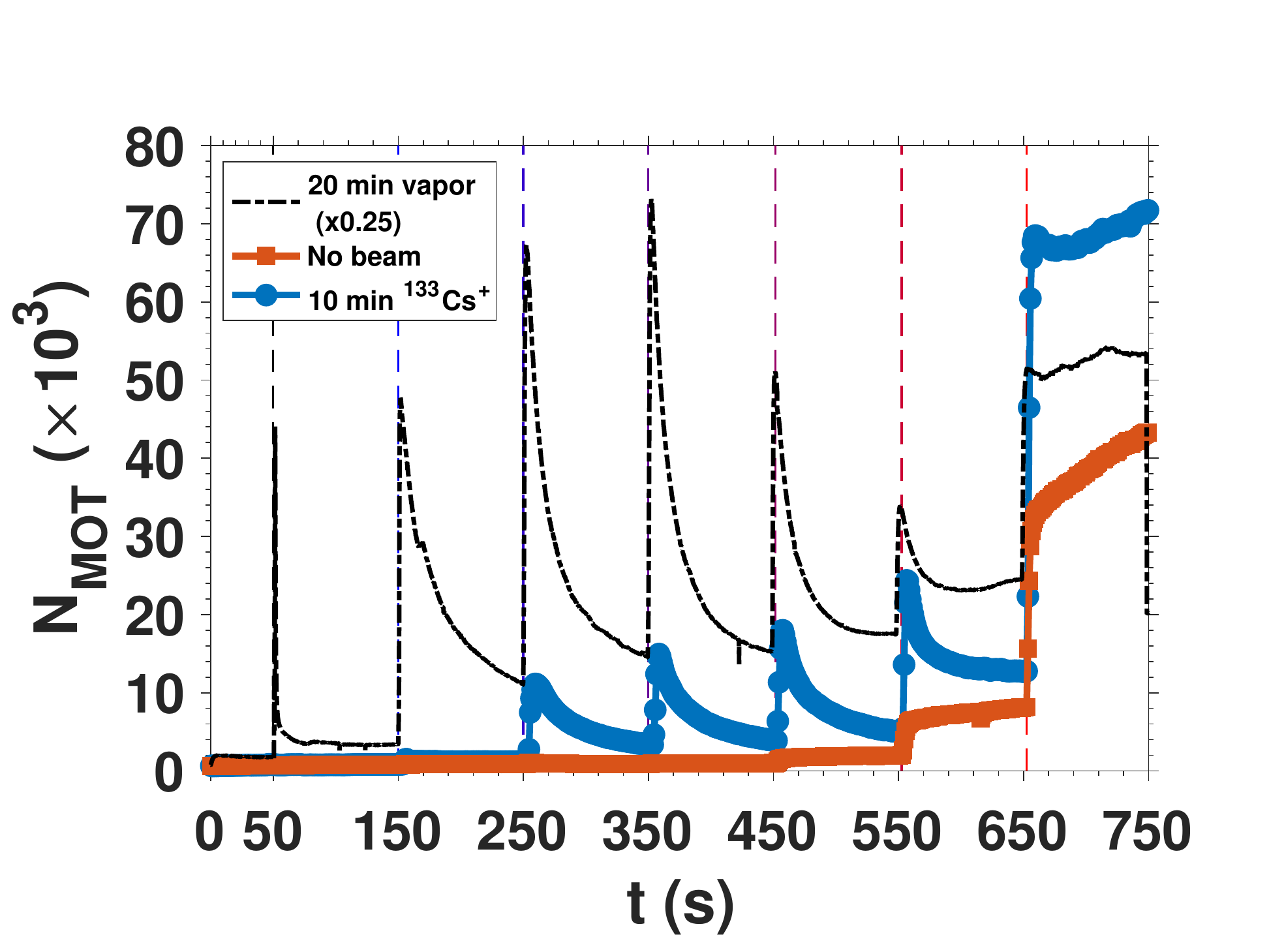}
\end{center}
\caption{(Color online) Different loading methods via Y. Thin black line: after \SI{20}{\minute} exposure to \SI{e10}{\per\centi\meter\cubed} (multiplied by 0.25). Red squares: Previously implanted beam. Blue disks: extra \SI{10}{\minute} of $^{133}$Cs$^{+}$ at \SI{1e-10}{\ampere} at the Y foil. The vertical dashed lines mark increase in T$_{\text{Y}}$. From left to right: $\text{T}_{\text{Y}}=\SI{691}{},\SI{767}{},\SI{832}{},\SI{887}{},\SI{931}{},\SI{972 }{},\SI{1010}{\kelvin}$.}
\label{fig:overview}
\end{figure}

Figure~\ref{fig:overview} shows N$_{\text{MOT}}$ as a function of time, when I$_{\text{DC}}$ is increased by \SI{0.5}{\ampere} between \SI{3.5}{\ampere}  (\SI{691}{\kelvin}) and \SI{6.5}{\ampere} (\SI{1010}{\kelvin})  every \SI{100}{\second}. The black thin trace shows the behavior when -- before the described sequence -- the Y at room temperature is exposed to a Cs thermal vapor \SI{e10}{\per\centi\meter\cubed} for \SI{20}{\minute}. In this case, atoms are adsorbed at the surface sites of the Y foil, and therefore are quickly ejected upon heating. This creates a spike in the MOT population, which is however rapidly reduced, given the finite density of superficially adsorbed atoms. Such rapid dynamics are typical of atoms adsorbed in surface sites \cite{jchemphyslnl, marmugi2014lpl}. Noticeably, the N$_{\text{MOT}}$ level reached after \SI{50}{\second} is higher than the initial level, thus indirectly confirming the quality of the PDMS coating: atoms can interact several times with the coated walls before being permanently lost because of adsorption or the vacuum system. The progressive decrease of the maximum peak at higher temperatures indicates that the main surface source of atoms is rapidly depleted. The behavior dramatically changes at high currents. In particular, at \SI{1010}{\kelvin} (I$_{\text{DC}}$=\SI{6.5}{\ampere}), a net increase of the atomic population and completely different loading dynamics are observed. The slow process is typical of bulk processes \cite{marmugi2014lpl}. 

% From 2017_12_11-133Cs_Systematics
% 02-a-NoBeam-Yincreased-500ms
% 05-a-After10mbeam-Yincreased-500ms
% 09-Vapour20m-Yincreased-50ms
% Calibration from 20180619 - Final fluorescence calibration.nb, rescaled for texposure.

When the vapor contribution to the Y is completely eliminated (red squares in Fig.~\ref{fig:overview}), the fast dynamics completely disappear. At high temperature, the bulk contribution due to previous implantations emerge again. In this condition, substantial MOT loading -- only with slow dynamics -- is observed only at high temperatures (T$_{\text{Y}}$=\SI{972}{\kelvin}, \SI{1010}{\kelvin}, corresponding to I$_{\text{DC}}$=\SI{6.0}{\ampere}, \SI{6.5}{\ampere}).

The same experiment is performed after further implantation of ``fresh'' $^{133}$Cs$^{+}$ ions, at an average current of \SI{e-10}{\ampere}. Remarkably, MOT loading can be clearly observed already with I$_{\text{DC}}$=\SI{4.5}{\ampere}, and the fast dynamics is again observed. At higher current, the dynamics progressively approach the slow regime attributed to the bulk contribution, although almost two times bigger than what was observed before the extra implantation. This is a direct evidence of the contribution -- both on the surface and bulk populations of the Y foil -- of the $^{133}$Cs$^{+}$ ionic beam to the MOT population.

\subsubsection{Loading from previous ion implantations}
\begin{figure}[htbp]
\begin{center}
\includegraphics[height=6.5cm]{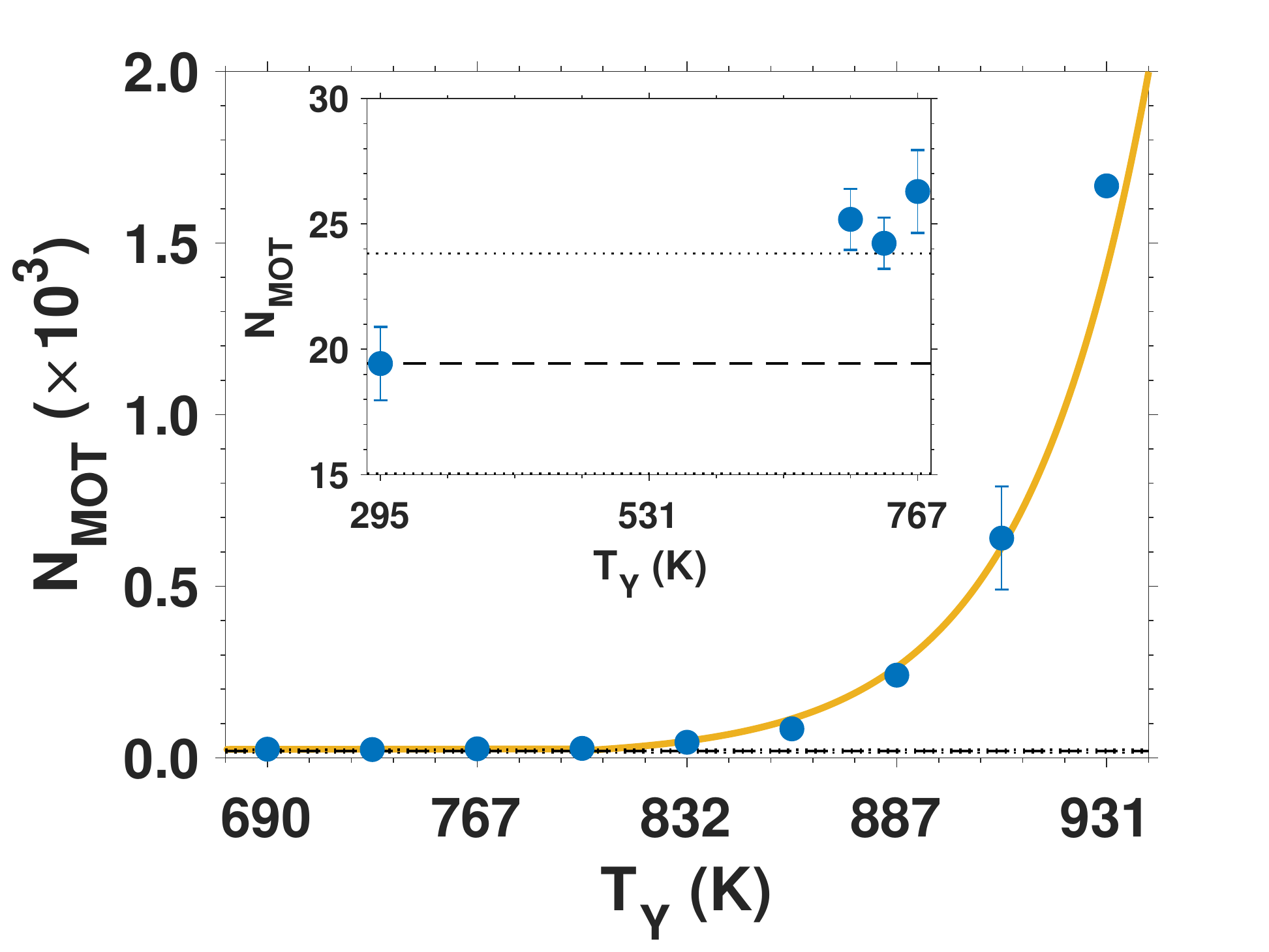}
\end{center}
\caption{(Color online) MOT loading from Cs atoms desorbed from Y, after previous $^{133}$Cs$^{+}$ implantations. N$_{\text{MOT}}$ vs T$_{\text{Y}}$. Dashed lines marks the background noise level, the dotted lines the 3$\sigma$ level. The yellow line is an exponential fit based on the Arrhenius equation applied to desorption. Inset: detail of the low-current regime, where values are consistent with no detection.}
\label{fig:lowcurrent}
\end{figure}

Figure~\ref{fig:lowcurrent} shows results concerning $^{133}$Cs MOT loading from previously implanted ions. It is noteworthy that in all cases there is no detected loading from background vapor.

% From 2018_04_20
% /Other 133Cs+ data/2018_04_20 - Systematics for low Y temp/Second effort-MOT optimised for low Y temp/Luca - New graphs
% Calibration from 20180619 - Final fluorescence calibration.nb, rescaled for texposure=1000 ms.

In this regime, desorption of neutral atoms from Y foil is produced only by thermal diffusion: previously implanted ions, once neutralized inside the foil, progressively diffuse out of the foil, with a clear exponential dependence on T$_{\text{Y}}$. This behavior is consistent with a purely thermal diffusion process. 

\begin{figure}[htbp]
\begin{center}
\includegraphics[height=6.5cm]{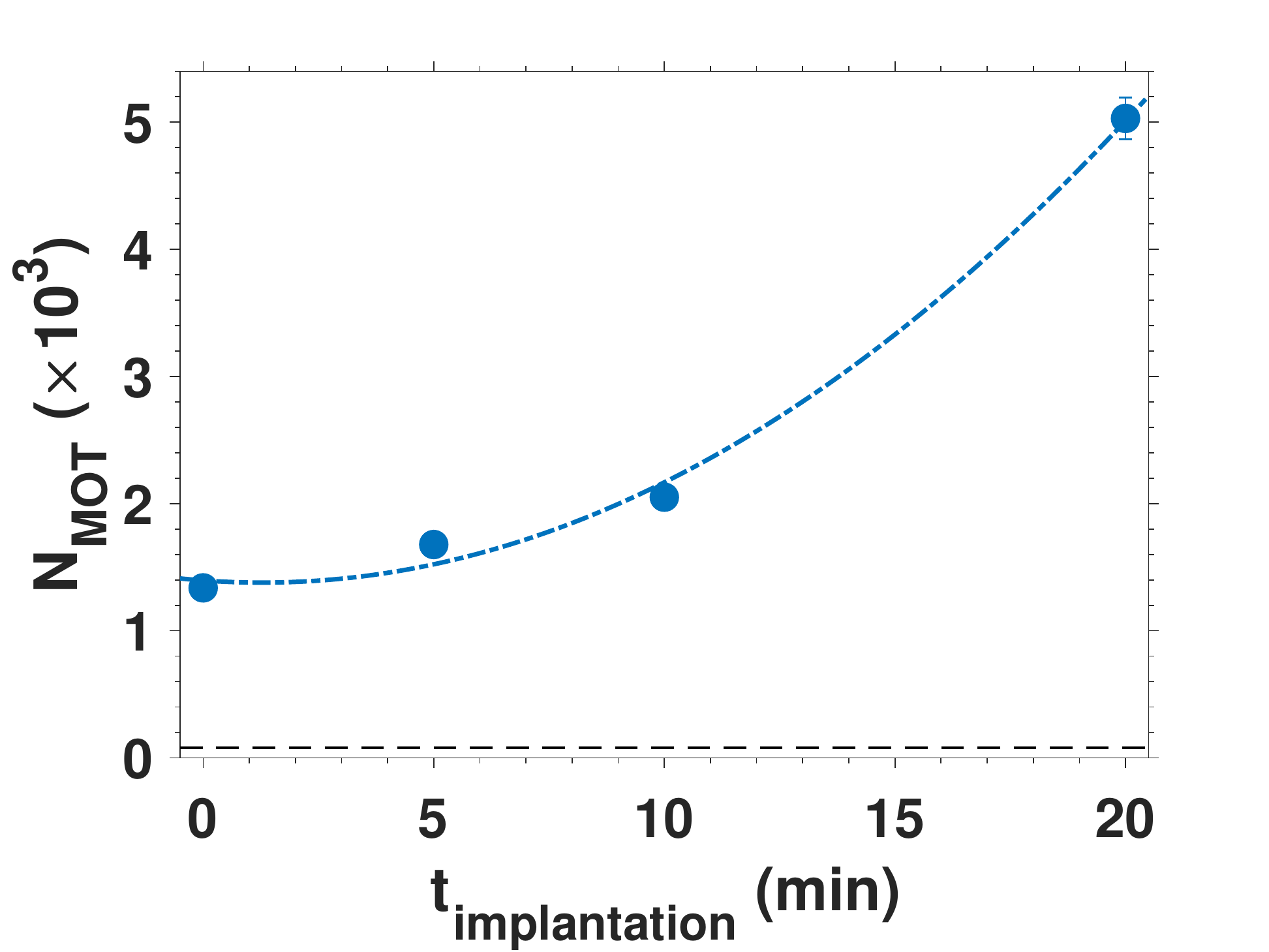}
\end{center}
\caption{(Color online) Loading exclusively from Cs atoms desorbed from Y, as a function of implantation duration (t$_{\text{implantation}}$) with \SI{e-10}{\ampere}. Y at 5.5 A. Quadratic fit is a guide for the eye. The dashed line marks the background noise level. Atoms trapped at t$_{\text{implantation}}=\SI{0}{\minute}$ were previously implanted. This level can be considered an offset when evaluating the impact of t$_{\text{implantation}}$.}
\label{fig:loading}
\end{figure} 
% From 2017_12_11
% Graphs in Finland/Beam Time Jyvaskyla - December 2017/Data Jyvaskyla/2017_12_11-133Cs_Systematics/06-e-MOT+20minbeam-Y5_5-1000ms
% Calibration from 20180619 - Final fluorescence calibration.nb, rescaled for texposure=1000 ms.

To confirm the gain in N$_{\text{MOT}}$ produced by the ion beam, we investigated the impact of previously implanted $^{133}$Cs$^{+}$ in Fig.~\ref{fig:loading}. Here, \SI{1e-10}{\ampere} of $^{133}$Cs$^{+}$ are implanted for a variable time t$_{\text{ion}}$, and the maximum population of the MOT is plotted against the implantation time (T$_{\text{Y}}$=\SI{930}{\kelvin}). The data confirms the expected growth in the number of trapped atoms for increasing implantation time.

\subsubsection{Direct loading from neutralized $^{133}$Cs$^{+}$ beam}
A second set of validation experiments was carried out to find the direct evidence of MOT loading from freshly implanted ions. In particular, the compatibility of the \SI{30}{\kilo\electronvolt} beam and the MOT is also assessed. This is an essential prerequisite for continuous implantation operation with radioactive beams.

\begin{figure}[b!]
\begin{center}
\includegraphics[height=6.5cm]{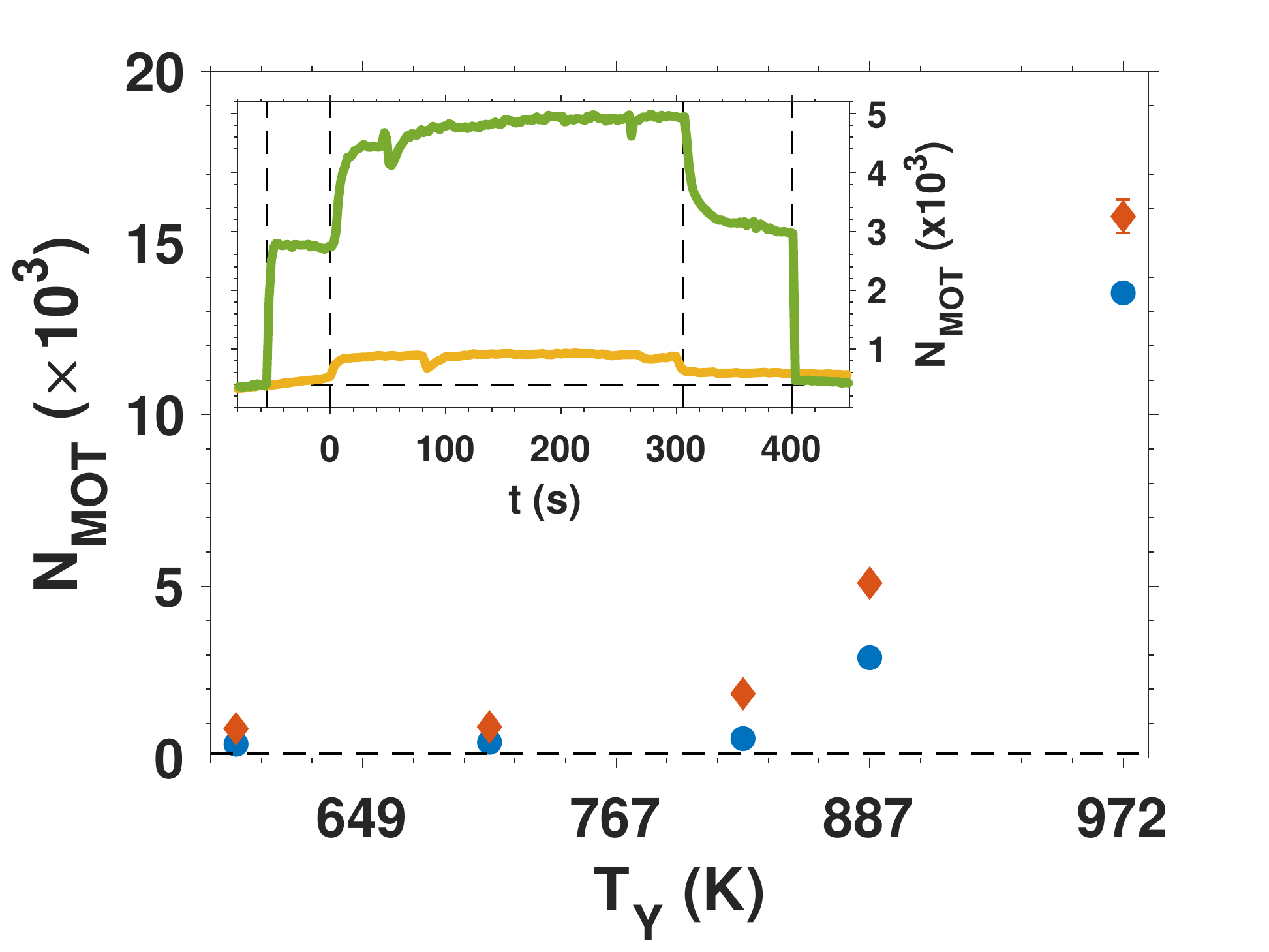}
\end{center}
\caption{\red{(Color online) Direct loading from neutralized} $^{133}$Cs$^{+}$ beam: N$_{\text{MOT}}$ before (blue disks) and after (red diamonds) exposure to \SI{e-10}{\ampere} $^{133}$Cs$^{+}$ beam, as a function of T$_{\text{Y}}$. Inset: MOT fluorescence measured at T$_{\text{Y}}=\SI{691}{\kelvin}$ and \SI{887}{\kelvin} (I$_{\text{DC}}=\SI{3.5}{\ampere}$ and \SI{5}{\ampere}). The vertical dashed lines mark the timestamps of magnetic field gradient on (t=-\SI{30}{\second}), ion beam on (t=\SI{0}{\second}), ion beam off (t=\SI{300}{\second}), magnetic field gradient off (t=\SI{400}{\second}). The horizontal dashed lines mark the detection threshold.}
\label{fig:all}
\end{figure} 
% From 2017_12_11 - MOT+beam
% 2017_12_11_17-06-55_Inset
% Calibration from 20180619 - Final fluorescence calibration.nb, rescaled for texposure. 

To this goal, a background MOT is loaded from previous implantations, by heating the Y foil. Once the population is stable, the $^{133}$Cs$^{+}$ beam is switched on for \SI{300}{\second}, and then switched off again. A typical result is shown in the inset of Fig.~\ref{fig:all}. Here, we note a significant increase of N$_{\text{MOT}}$ both with initial zero population (yellow trace) and with initial substantial population (green trace). The latter also confirms the compatibility of the MOT and the ion beam. In both cases, the response of the MOT population to the incoming beam is observed in around \SI{5}{\second} after the ion beam is switched on. Noticeably, after the beam is switched off, the MOT population decays to a larger level than the initial one.

There is a relatively small and constant increase in N$_{\text{MOT}}$ when the ion beam is on for T$_{\text{Y}}\leq\SI{767}{\kelvin}$ (I$_{\text{DC}}\leq\SI{4}{\ampere}$). On the contrary, from T$_{\text{Y}}\geq\SI{832}{\kelvin}$ (I$_{\text{DC}}\geq\SI{4.5}{\ampere}$), a substantial increase of the MOT population is observed, which increases with the foil temperature. This is consistent with thermally activated diffusion of neutralized atoms. \red{The role of thermal diffusion is further confirmed by the larger increase in the MOT population observed in the inset of Fig.~\ref{fig:all} with increasing T$_{\text{Y}}$: this effect is attributed to more effective desorption of ions implanted in deeper regions of the foil, aided by the higher temperature.}

\begin{figure}[htbp]
\begin{center}
\includegraphics[height=6.5cm]{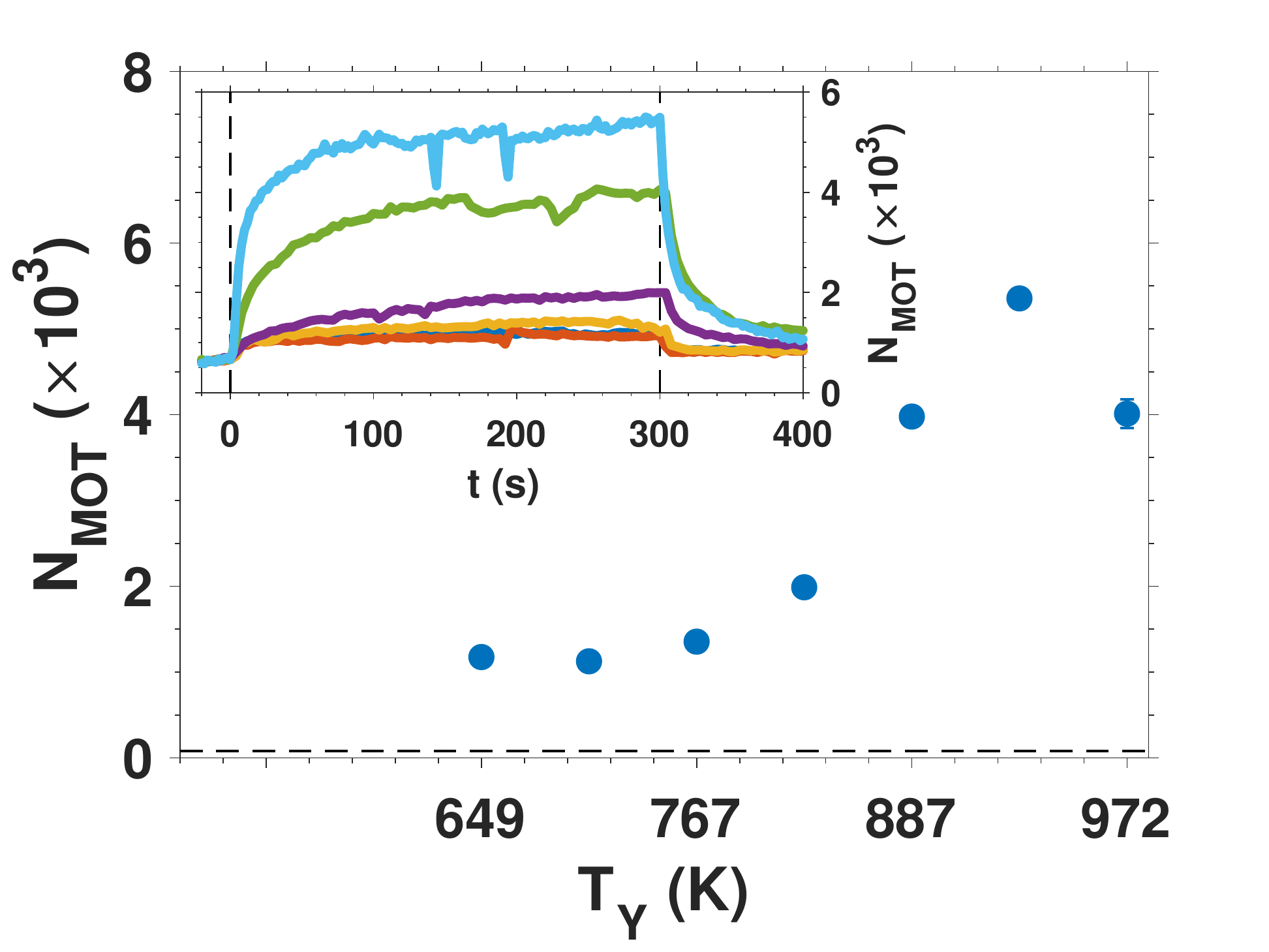}
\end{center}
\caption{(Color online) Direct contribution from the Y foil. Maxima in N$_{\text{MOT}}$ versus Y temperature. \SI{e-10}{\ampere} of $^{133}$Cs$^{+}$. Note effect of sputtering at T$_{\text{Y}}\leq\SI{767}{\kelvin}$. Inset: real-time fluorescence evolution at different Y temperatures (from bottom to top: T$_{\text{Y}}=\SI{649}{}, \SI{691}{}, \SI{767}{}, \SI{831}{}, \SI{887}{}, \SI{929}{K}$).}
\label{fig:beamonly} 
\end{figure}

This regime  corresponds to the most direct evidence of operation of the complex ion beam/neutralization/organic coating/laser trapping. Therefore, it is also tested when no loading from vapor or the Y foil is observed (i.e. when there is no MOT unless new ions are implanted and neutralized). This mimics the working conditions with radioactive atoms, where the instability of the species of interest can make ``off-line'' loading practically impossible. In Fig.~\ref{fig:beamonly}, such a regime is demonstrated and characterized. Evidence of direct loading from the ion beam (on at t=\SI{0}{\second} and off at t=\SI{300}{\second}) is observed at any temperature between \SI{607}{\kelvin} and \SI{972}{\kelvin} (I$_{\text{DC}}$ between \SI{3.0}{\ampere} to \SI{6.0}{\ampere}). A temperature-independent increase of N$_{\text{MOT}}$ is also observed at T$_{\text{Y}}\leq\SI{767}{\kelvin}$ (I$_{\text{DC}}\leq\SI{4}{\ampere}$).  \red{We also observed this effect well below 400 K, without any obvious thresholds in temperature.}

% From 2017_12_11 - BeamOnly
% 2017_12_11_MOTmean_comparison 
% 2017_12_11_MOTmean_comparison-AVERAGES
% Final graph: 2017_12_11_MOTmean_comparison_all.fig
% Calibration from 20180619 - Final fluorescence calibration.nb, rescaled for texposure. 

Such an increase at low T$_{\text{Y}}$ -- independent from the Y heater current and also observed in Fig.~\ref{fig:all} -- can be explained with direct ejection of previously neutralized atoms at the surface of the foil by the high-energy incoming ions. This effect, practically similar to sputtering, can be directly observed here because of the high kinetic energy of the incoming ions (\SI{30}{\kilo\electronvolt}).

\red{With the currently available diagnostic on the ion beam in the science chamber (see also Sec.~\ref{sec:outlook}), we did not observe any increase in MOT losses due to the presence of the incoming ions. This can be also attributed to the small vertical offset between the MOT center and the ion beam core, as determined by optimizing the position of the trap for maximum population.}

\red{Notwithstanding the currently limited direct information on the number of incoming ions and on the ``freshly'' desorbed $^{133}$Cs atoms (due to the bulk contribution), a lower bound for the trapping efficiency based on the results shown in Fig.~\ref{fig:beamonly} can be estimated. By using \SI{6e8}{\text{atoms}\per\second} (i.e. \SI{e-10}{\ampere} of $^{133}$Cs$^{+}$), and an average MOT lifetime of $\tau=\SI{2.85}{\second}$ (1/e time, measured in a separate experiment), the overall efficiency at T$_{\text{Y}}=\SI{929}{\kelvin}$ is:  $\eta \geq \SI{3e-6}{}$.}

\subsection{Implantation test runs}\label{subsec:radioactive}
\begin{figure*}[htpb]
\begin{center}
\includegraphics[height=6.5cm]{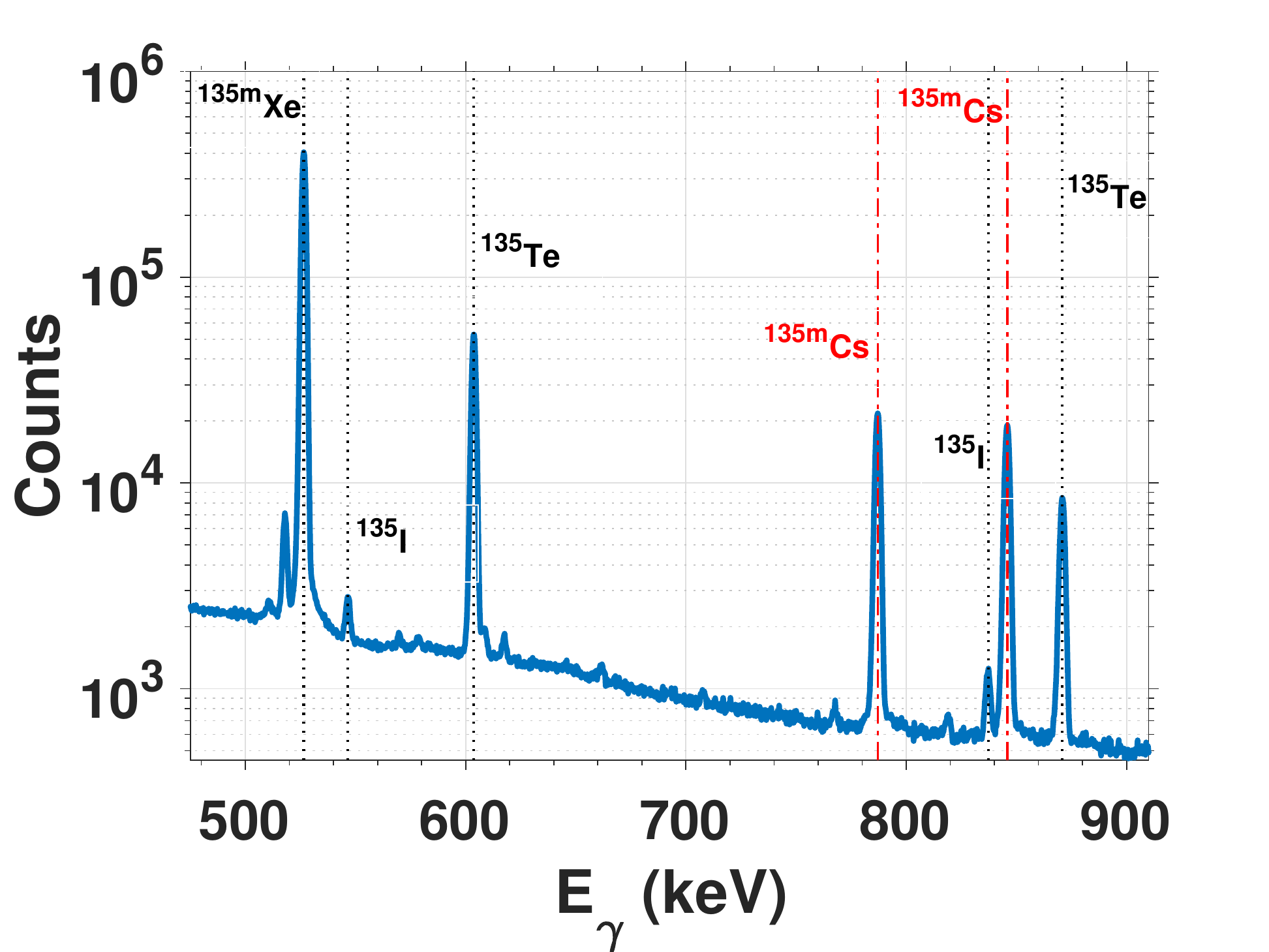}
\includegraphics[height=6.5cm]{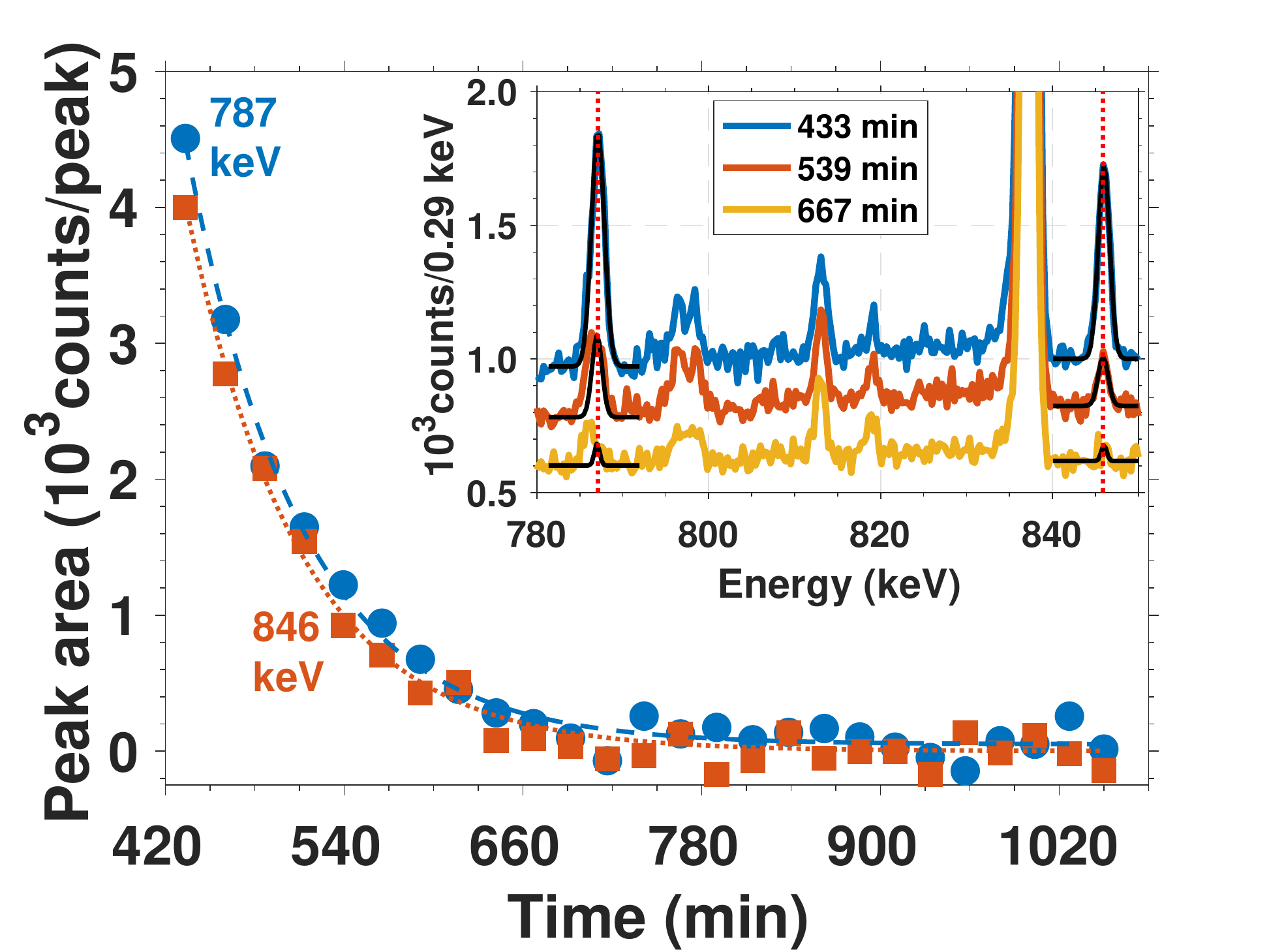}
\caption{Beam spectra with mass number A=135, after proton-induced fission of a U target and extraction. Left: $\gamma$ spectrum produced by a Ge detector after \SI{30}{\minute} implantation. $^{135m}$Cs implantation rate is \SI{1.5e4}{\per\second}. Right: Off-line decay curve of both $\gamma$ transitions of $^{135m}$Cs, obtained after implantation in the Y foil. Inset: $\gamma$ spectra of the implanted foil at selected timestamps after implantation. The $^{135m}$Cs $\gamma$ peaks are highlighted by black fitting curves. \label{fig:detection}}
\end{center}
\end{figure*}
% From May beamtime. 

Lastly, we report on the implantation test runs conducted with a $^{135,135m}$Cs$^{+}$ beam. This allowed us to validate the production, transport, mass selection, and implantation procedure with one of the species of interest \cite{marmugi2018}. $^{135m}$Cs (J$^{\pi}=19/2^{-}$) has a half-life of T$_{1/2}=\SI{53}{\minute}$, before decaying via an M4 (E$_{\gamma}=\SI{846.1}{\kilo\electronvolt}$) and an E2 (E$_{\gamma}=\SI{786.8}{\kilo\electronvolt}$) $\gamma$ transition to the ground state $^{135}$Cs (J$^{\pi}=7/2^{+}$, T$_{1/2}=\SI{2.3e6}{\text{yrs}}$). Because of these characteristics, it is well-suited for implantation tests: it can be directly detected in the ion beam, or it can be monitored after implantation to observe the $\gamma$ decay.

During this test, both the ground- and the isomeric state of $^{135}$Cs are populated by proton-induced fission of natural U using a primary beam intensity up to \SI{6}{\micro\ampere}. To verify the production of the isomeric state, the mass separated A/q=135 beam is routed towards the spectroscopy line located to the right of the electrostatic switchyard. The ion beam is then implanted for \SI{30}{\minute} into a thin foil mounted at the end of the beam line which is directly viewed from behind by a Ge detector (Canberra GC7020, energy range from \SI{40}{\kilo\electronvolt} to \SI{10}{\mega\electronvolt}) in order to detect the emitted $\gamma$ photons. In Fig.~\ref{fig:detection}, we show the ``on-line'' $\gamma$ spectra measured with the Ge detector. Clear signatures of the M4 and E2 $\gamma$ emissions are observed. In addition to detection during implantation, $\gamma$ emissions are also monitored during a \SI{30}{\minute} decay period when no beam is implanted. Following analysis of the \SI{787}{\kilo\electronvolt} and \SI{846}{\kilo\electronvolt} peaks -- taking into account the solid angle and detector efficiency -- an implantation rate of \SI{2e4}{\per\second} was detected, with \SI{5}{\micro\ampere} intensity of the primary beam.

A transmission efficiency of $\sim$70$\%$ has been measured from the switchyard to the end of the spectroscopy line via implantation of A/q=112 fission products into foils mounted directly in front of Si charged-particle detectors. In this case, $^{112}$Rh (T$_{1/2}=\SI{6.8}{\second}$) is used for beam tuning due to the suitable half-life. To verify the corresponding value from the switchyard to the neutralizer foil in the science chamber, a trace amount of stable $^{136}$Xe gas is added to the He buffer gas. The Xe is ionized during on-line operation of the gas cell and can be detected in quantities suitable for Faraday cups. During the on-line production of $^{135m}$Cs, \SI{95}{\nano\ampere} of $^{136}$Xe was detected in the switchyard. Of this, \SI{38}{\nano\ampere} was measured at the Y neutralizer foil, thus leading to a transmission efficiency of $\sim$40\%. On the basis of these numbers, it is therefore estimated that \SI{1.2e4}{\per\second} $^{135m}$Cs were implanted into the neutralizer foil.

$^{135}$Xe $\gamma$ signatures are also detected (Fig.~\ref{fig:detection}a)), as well as those of other by-products of the fission reaction. We note that this will not represent a problem for the purity of the cold atomic samples: further selection -- beyond the charge-to-mass ratio 135 -- will be automatically performed by the laser cooling process. Unwanted species (or ground state, as opposed to the isomeric state) will not be trapped.

To confirm successful implantation of $^{135m}$Cs, in Fig.~\ref{fig:detection}, we show the ``off-line'' $\gamma$ spectra, measured by extracting the Y foil from the vacuum chamber after \SI{2}{\day} implantation at \SI{3e-14}{\ampere} $^{135m}$Cs$^{+}$. Data are obtained in a low-background off-line chamber, at regular time intervals, starting from \SI{7}{\hour} after the conclusion of implantation. This time was necessary to allow safe transfer of the freshly implanted Y foil to the low-background chamber, due to the shorter-living activity of other A/q=135 species which was implanted in addition to Cs. The counting station is equipped with a Canberra BE3825 detector, with energy range from \SI{3}{\kilo\electronvolt} to \SI{3}{\mega\electronvolt}. The detector head is housed in a lead castle with N$_{2}$ atmosphere to reduce the background counts.

The decrease of the two $\gamma$ emission peaks of $^{135m}$Cs can be clearly observed as a function of time. Interfering emission at \SI{785}{\kilo\electronvolt} is also detected. This is attributed to the products of the $^{135}$I $\beta$-decay: the  \SI{785}{\kilo\electronvolt} photon is produced by the fast de-excitation of the daughter nucleus $^{135}$Xe \cite{nucleardb}. The longer half-life (T$_{1/2}$=\SI{6.58}{\hour}) of  $^{135}$I allows one to discriminate and remove its contribution. By combining together data from the two $\gamma$ peaks typical of the spectrum of $^{135m}$Cs, we measured a half-life of T$^{\text{exp}}_{1/2}$=\SI{53.5\pm1.4}{\minute}. The uncertainty in this case is purely statistical. The measured half-life is consistent with the previously reported value T$_{1/2}=\SI{53\pm2}{\minute}$ \cite{halflife}. This confirms the identification of $^{135m}$Cs.

\section{Conclusions and Outlook}\label{sec:outlook}
We have designed, built, and tested an experimental facility for laser cooling and trapping of Cs isotopes and isomers. Production of the desired radioactive species is achieved either by proton-induced fission of U targets, or by a fusion-evaporation reaction.  Nuclei are extracted as +1 ions from the target area, electrostatically accelerated, and mass separated in the IGISOL facility. Ions are implanted in a thin Y foil, where they dissipate their kinetic energy, acquire an extra electron (``neutralization''), and are stored before undergoing thermal diffusion to the experimental chamber. Here, a PDMS organic coating allows efficient laser trapping. Tests conducted with a local beam of $^{133}$Cs$^{+}$ confirmed the correct operation of the system, and highlighted the direct contribution of the ion beam to the population of a magneto-optical trap. Atoms are cooled down to \SI{150}{\micro\kelvin}, with a density of \SI{e10}{\per\centi\metre\cubed}. \red{Preliminary estimates with $^{133}$Cs$^{+}$ identified a lower limit for the trapping efficiency of $\eta\geq3\times10^{-6}$. This is compatible with successful experiments with other radioactive species \cite{gwinner1994}, with potential for further improvement \cite{efficiency}.} This opens up new scenarios and possibilities, from high-precision laser spectroscopy of unknown electronic states for investigations on the isomeric charge radius anomaly, to nuclear forensics. The experiment also constitutes the building block of a long-term endeavor for multi-body physics in ultra-cold isomeric gases. 

In the immediate future, the $^{133}$Cs MOT will be investigated over a wider range of implantation rates, following the installation of new particle diagnostics in the beam line. This will allow us to extend testing to lower ionic currents and to measure the overall efficiency of the setup. Ultimately, this will identify the lower trapping limit as a function of the number of implanted ions. This information is paramount to tailor the most effective approach for laser cooling and trapping of radioactive species.

 \section*{Acknowledgements}
This work was partially funded by the H2020-EU.1.3.2 programme through the Marie Curie Fellowship 2020-MSCA-IF-2014 ``GAMMALAS'' to L.~M. (Proj. Ref. 657188), and by the Royal Society (IE151199). P.~M.~W. acknowledges STFC support under grant no. ST/L005743/1.\\
We thank Krishna Jadeja (UCL) and Yuval Cohen (UCL), who recently joined this project after completion of the setup.

\section*{Declarations of Interest}
Declaration of interest: none.

\section*{References}
%\bibliography{cesium.bib}
%\bibliographystyle{elsarticle-num}

\end{document}